\documentclass[preprintnumbers,superscriptaddress,showkeys,showpacs,byrevtex]{revtex4}
\usepackage{amsmath,amsfonts,amssymb,amscd,amsxtra,amsthm}
\usepackage{graphicx}
\usepackage{epstopdf}
\usepackage{bm}
\usepackage{feynmp}
\usepackage{xcolor}
\usepackage{feynmp-auto}
\usepackage{amsmath,amsfonts,amssymb,amscd,amsxtra,amsthm}
\begin{document}
\preprint{PKNU-NuHaTh-2021-03}
\title{Studies on the $K^*\Sigma$ bound-state via $K^+p\to K^+\phi\,p$}
%--------------------------------------------------
\author{Seung-il Nam}
\email[E-mail: ]{sinam@pknu.ac.kr}
\affiliation{Department of Physics, Pukyong National University (PKNU), Busan 48513, Korea}
\affiliation{Center for Extreme Nuclear Matters (CENuM), Korea University, Seoul 02841, Korea}
\affiliation{Asia Pacific Center for Theoretical Physics (APCTP), Pohang 37673, Korea}
%--------------------------------------------------
\date{\today}
\begin{abstract}
In the present work, we investigate the hidden-strangeness production process in the $S=+1$ channel via $K^+p\to K^+\phi\,p$, focussing on the exotic \textit{pentaquark} molecular $K^*\Sigma$ bound state, assigned by $P^+_s(2071,3/2^-)$. For this purpose, we employ the effective Lagrangian approach in the tree-level Born approximation. Using the experimental and theoretical inputs for the exotic state and for the ground-state hadron interactions, the numerical results show a small but obvious peak structure from $P^+_s$ with the signal-to-background ratio $\approx1.7\,\%$, and it is enhanced in the backward-scattering region of the outgoing $K^+$ in the center-of-mass frame. We also find that the contribution from the $K^*(1680,1^-)$ meson  plays an important role to reproduce the data. The proton-spin polarizations are taken into account to find a way to reduce the background. The effects of the possible $27$-plet pentaquark $\Theta^{++}_{27}$ is discussed as well. 
\end{abstract}
\maketitle
%--------------------------------------------------
\section{Introduction}
%--------------------------------------------------
Hadronic interactions have been one of the most important subjects in the strongly-interacting systems governed by quantum chromodynamics (QCD). From the interactions, various hadronic states can be constructed.  The most interesting hadronic states must be exotic hadrons, which are beyond the minimal meson and baryon configurations, i.e., $q\bar{q}$ and $qqq$ in their color singlets: Tetraquarks, pentaquarks, hadronic moleculars, di-baryons, and so on.  {However, all of these exotics have not been fully confirmed yet experimentally. For instance, although QCD does not seem to prohibit a $qqqq\bar{q}$ configuration, the existence of light pentaquark resonance states, including the famous $\Theta^+$ pentaquark baryon~\cite{Diakonov:1997mm,Nam:2003uf,Nakano:2008ee,Yao:2006px}, has been unsettled, while various hidden-charm heavy pentaquark molecular bound states were observed in the $J/\psi$-$p$ invariant mass from the heavy baryon decay $\Lambda^+_b\to K^-J/\psi\,p$ at LHC$_b$, $P^+_c(4312,4440,4457)$~\cite{Aaij:2019vzc}. Interestingly, those heavy-pentaquark states were not measured in the $J/\psi$ photoproduction off the proton target in the GlueX experiment of CLAS at the Jefferson laboratory~\cite{Ali:2019lzf}, indicating the possible smaller photon couplings of the states. }

We have the following motivations for the present work: As mentioned above, the pentaquark molecular $\bar{D}^*\Sigma_c$ bound-state $P^+_c(4457)$ was found experimentally from the decay with a \textit{hidden-charm} vector meson~\cite{Aaij:2019vzc}: $P^+_c[\bar{D}^*\Sigma_c]\to J/\psi[c\bar{c}]\,p$.  Considering an analogous mechanism for the light-flavor sector and assuming that the hidden-flavor channel is the key to observe the pentaquark bound state, there can be a $K^*\Sigma$ bound state appearing in the invariant mass of the \textit{hidden-strange} vector meson $\phi(1020)$ and proton, being assigned by $P^+_s$ for convenience: $P^+_s[K^*\Sigma]\to \phi[s\bar{s}]\,p$.  Interestingly, from various experimental data~\cite{Mibe:2005er,Mizutani:2017wpg,Dey:2014tfa}, a bump-like structure is observed at $\sqrt{s}=(2.0\sim2.1)$ GeV for the $\phi$-meson photoproduction off the proton target, although the origin of the bump is not fully understood~\cite{Kiswandhi:2010ub,Kiswandhi:2016cav}.  Moreover, from theories, the $K^*\Sigma$ molecular bound state was suggested from the coupled-channel approach with the hidden-local symmetry and turned out to couple to the $\phi\,p$ channel rather strongly~\cite{Khemchandani:2011et}. Also in Refs.~\cite{Oset:2009vf,Ramos:2013wua}, the bound state was investigated in a similar approach and turns out to be crucial around $\sqrt{s}\lesssim2.0$ GeV. Hence, we assign this molecular bound state  to $P^+_s$, whose mass and spin-parity will be taken from theories, and it can be viewed as the light-flavor partner of $P^+_c(4457)$, {although the spin-parity quantum numbers of the heavy pentaquarks including $P^+_c(4312,4440)$ have not been uniquely determined yet.}

Considering the motivations given above, we would like to study the hidden-strangeness production process with the $S=+1$ kaon beam off the proton target, i.e., $K^+p\to (K^+P^+_s) \to K^+\phi\,p$. For this purpose, we employ the effective Lagrangian method at the tree-level Born approximation. As for the interaction structures, we basically make use of the pseudo-scalar(PS)-meson--baryon Yukawa interactions for the baryon-intermediate $s$- and $u$-channel diagrams, in addition to the $\phi$-meson exchange $t$-channel diagrams. As shown in the LHC$_b$ experiment for $B^+\to J/\psi\phi K^+$~\cite{Aaij:2016iza}, the contribution from $K^*(1680,1^-)$, which decays into $K\phi$, is also taken into account. A possible $27$-plet pentaquark state $\Theta^{++}_{27}$ is included as well to verify the effects of its existence to physical observables. The relevant interaction strengths are taken from the well-known Nijmegen potential model~\cite{Rijken:1998yy} and the coupled-channel method with the hidden local symmetry~\cite{Khemchandani:2011et}. Phenomenological form factors are also taken into account, and the cutoff masses for the form factors determined to reproduce the presently available experimental data for $K^+p\to K^+\phi\,p$~\cite{Flaminio:1983fw,Bishop:1970at,Moser:1977mp}. 

From the numerical results, we observe the strong enhancements of the total cross section with the $K^*(1680)$ contribution as the center-of-mass (cm) energy increases, due to the higher momentum-dependent Lorentz structure of the $KK^*\phi$ interaction vertex. In the Dalitz plot analyses, we find the band structure with $\Gamma=14$ MeV for $P^+_s$, and the $P^+_s$ contribution interferes with the background constructively. As the energy increases larger than $\sqrt{s}\approx2.65$ GeV, the $K^*(1680)$ contribution dominates the cross section.  In the $\phi\,p$-invariant-mass plots,  we find an obvious peak structure from $P^+_s$ with the signal-to-background ratio $\approx1.7\%$ at $\sqrt{s}=2.65$ GeV. There is the considerable cross-section enhancement from the $K^*(1680)$ contribution for $M(\phi p)\lesssim2.05$  GeV in the $\phi\,p$-invariant-mass plots as expected from the total cross section. There appears a broad $K^*(1680)$ peak for the higher cm energy in the $K^+\phi $-invariant-mass plots. In addition to the $P^+_s$ and $K^*(1680)$ contributions, the backgrounds with the $\Lambda(1115)$-hyperon intermediate states with the $\phi$ meson emitted from the kaon beam turns out out be the most important and largest contribution to reproduce the cross section data.  

A method to reduce the background is suggested for the better signals of $K^*(1680)$ and $P^+_s$, using the different initial- and final-state proton-spin combinations. When the two spins are opposite to each other, the background is largely suppressed, since the spin non-flip process only survives in the scattering amplitude. As for the opposite case, the $K^*(1680)$ peak manifests itself in the $K^+\phi $-invariant-mass  plots. In contrast, the $P^+_s$ peak is rather dubious for the opposite case in the $\phi\,p$-invariant-mass plots, since $P^+_s$ strongly interferes with the background  constructively, not with $K^*(1680)$. The angular dependence shows mild backward-scattering enhancements as functions of the outgoing $K^+$ angle in the cm frame, and this behavior was originated from the $u$-channel propagators in the scattering  amplitudes for the $K^*(1680)$ and background contributions. Therefore, the signal of $P^+_s$ is also amplified in the backward-scattering region.  Finally, we investigate the possible contribution from the $27$-plet pentaquark $\Theta^{++}_{27}$. Tuning the coupling strength of the pentaquark, we observe a diagonal band structure from the $\Theta^{++}_{27}$ in the Dalitz plot, and it starts to interfere strongly with $K^*(1680)$ for $M(\phi p)\gtrsim2.05$ GeV.

The present paper is organized as follows: The theoretical framework is briefly introduced in Section II. The Section III is devoted to the numerical results with detailed discussions. The summary is given in the final Section. 

%--------------------------------------------------
\section{Theoretical Framework}
%--------------------------------------------------
%FIGURE
\begin{figure}[t]
\includegraphics[width=16cm]{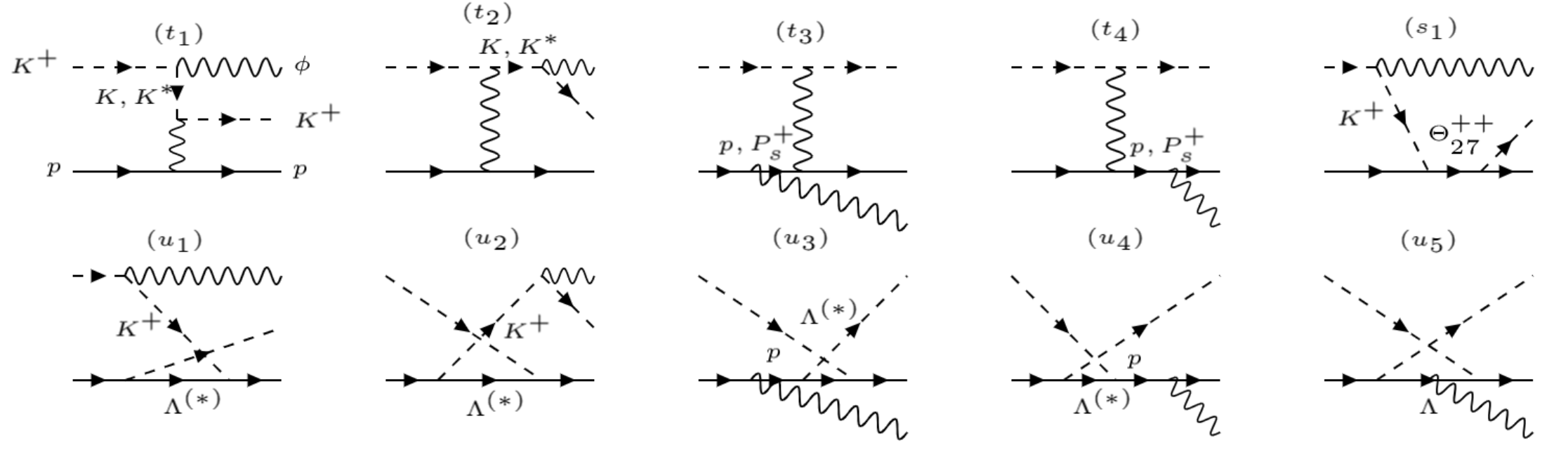}
\caption{Relevant tree-level Feynman diagrams for $K^+p\to K^+\phi\,p$. The dashed, solid, and wavy lines indicate the meson, baryon, and $\phi(1020)$ meson, respectively. The four momenta of the initial state $(K^+,p)$ and final state $(K^+,\phi,p)$ are defined by $(k_1,k_2)$ and $(k_3,k_4,k_5)$, respectively. For convenience, we separate the diagrams into three categories: The $\phi$-meson exchange diagrams in the $t$ channel ($t_{1\sim4})$, the $\Theta^{++}_{27}$-pole diagram in the $s$ channel ($s_{1}$), and the $\Lambda^{(*)}$-intermediate diagrams in the $u$ channel ($u_{1\sim5}$). The off-shell strange mesons can be $K(495,0^-)$ for the $(t,u,s)$ and $K^*(1680,1^-)$ for the $t_{1,2}$ channel, whereas the off-shell baryons for the $t_{1\sim4}$ and $u_{1\sim5}$ channels are $(p,P^+_s)$ and $(p,\Lambda^{(*)})$.}
\label{FIG0}
\end{figure}
%FIGURE

In this Section, we provide the theoretical framework to investigate the $K^+p\to K^+\phi\,p$ reaction process. In Fig.~\ref{FIG0}, we depict the relevant tree-level Feynman diagrams for it. The dashed, solid, and wavy lines indicate the meson, baryon, and $\phi(1020)$ meson, respectively. The four momenta of the initial state $(K^+,p)$ and final state $(K^+,\phi,p)$ are defined by $(k_1,k_2)$ and $(k_3,k_4,k_5)$, respectively. For convenience, we separate the diagrams into three categories: The $\phi$-meson exchange diagrams in the $t$ channel ($t_{1\sim4})$, the $\Theta^{++}_{27}$-pole diagram in the $s$ channel ($s_{1}$), and the $\Lambda^{(*)}$-intermediate diagrams in the $u$ channel ($u_{1\sim5}$). The off-shell strange mesons can be $K(495,0^-)$ for the $(t,u,s)$ and $K^*(1680,1^-)$ for the $t_{1,2}$ channel, whereas the off-shell baryons for the $t_{1\sim4}$ and $u_{1\sim5}$ channels are $(p,P^+_s)$ and $(p,\Lambda^{(*)})$. 

As for $P^+_s$ as the pentaquark molecular $K^*\Sigma$ bound state, we employ the theoretical results from Ref.~\cite{Khemchandani:2011et}, in which the hidden-local symmetry was taken into account with the coupled-channel Bethe-Salpeter equation, resulting in $P^+_s(2071,3/2^-)$ as the isospin $1/2$ pole, with its full decay width $\Gamma_{P^+_s}=14$ MeV. Note that in Ref.~\cite{Gao:2017hya}, the $s$-wave exotic bound state was estimated with $J^P=3/2^-$ nucleon resonance with its mass $\sim2064$ GeV via the quark delocalization color screening model (QDCSM). We consider the ground state $\Lambda(1115,1/2^+)$, $\Lambda(1405,1/2^-)$, and $\Lambda(1520,3/2^-)$ for the $\Lambda$-hyperon contributions. We, however, take only the ground-state one for the diagram $(u_5)$ into account, since little information is available for the $\phi\Lambda^*\Lambda^*$ vertex. Similarly, we do not include the proton resonances for brevity. 

In the Review of Particle Physics~\cite{Tanabashi:2018oca}, there are several strange mesons, decaying into $K^+\phi$, such as $K_1(1650,1^+)$ and $K^*(1680,1^-)$, although their experimental confirmations are still poor. Nonetheless, as shown in the LHC$_b$ experiment for $B^+\to J/\psi\phi K^+$~\cite{Aaij:2016iza}, among the various strange mesons, in the vicinity of the invariant mass $M(K^+\phi)=1.7$ GeV, the $K^*(1680,1^-)$ contribution is the most dominant one, in addition to the non-resonant background, whereas the $K_1(1650,1^+)$ contribution seems appearing in the higher-mass region $M(K^+\phi)\approx 1.8$ GeV. Note that our purpose of the present work is to study the $P^+_s$ bound state, which appears in the vicinity of $M(\phi\,p)=2$ GeV, and this mass region corresponds to $M(K^+\phi)=(1.6\sim1.7)$ GeV. Hence, we exclude the $K_1(1650,1^+)$ contribution rather safely for the numerical calculations and to reduce theoretical uncertainties. 

As shown in the diagram $(s_1)$ of Fig.~\ref{FIG0}, there can be a contribution from the baryonic exotic state, such as the $27$-plet pentaquark baryon $\Theta^{++}_{27}$. Its physical property was studied in Ref.~\cite{Wu:2003xc}, and its mass, spin-parity, and full width turn out to be $1.60$ GeV, $3/2^+$, and $\lesssim43$ MeV, respectively, via the chiral soliton model (ChSM).  From the decay width we obtain $g_{KN\Theta^{++}_{27}}\lesssim2.06$. 

To compute the invariant amplitudes for the diagrams in Fig.~\ref{FIG0}, the effective Lagrangians for the relevant interaction vertices are defined as follows:
%EQUATION>>>
\begin{eqnarray}
\label{eq:LAG}
{\cal{L}}_{VPP}&=&ig_{VPP}V^{\mu}\left[(\partial_{\mu}P^{\dagger})P-(\partial_{\mu}P)P^{\dagger}\right],
\cr
\mathcal{L}_{VVP}&=&\frac{g_{VVP}}{M_P}\epsilon^{\mu\nu\sigma\rho}F_V^{\mu\nu}F^{\sigma\rho}_VP+\mathrm{h.c.},
\cr
\mathcal{L}_{PNB_{1\pm}}&=&ig_{PNB_{1\pm}}\bar{N}
P\Gamma^\pm B_{1\pm}+\mathrm{h.c.},
\cr
\mathcal{L}_{PNB_{3\pm}}&=&\frac{g_{PNB_{3\pm}}}{M_P}
\bar{N}
(\partial_{\mu}P)\gamma_5\Gamma^\pm B^{\mu}_{3\pm}+\mathrm{h.c.},
\cr
\mathcal{L}_{VNB_{1\pm}}&=&g_{VNB_{1\pm}}
\bar{N}\Gamma^\pm\gamma_5 V_{\mu}\gamma^{\mu}B_{1\pm}+\mathrm{h.c.},
\cr
\mathcal{L}_{VNB_{3\pm}}&=&\frac{ig_{VNB_{3\pm}}}
{M_V}\bar{N}\Gamma^\pm F_{V\mu\nu}\gamma^{\nu}B^\mu_{3\pm}+\mathrm{h.c.},
\end{eqnarray}
%EQUATION>>>
where $P$ and $V$ designate the fields for the pseudo-scalar and vector mesons, whereas $N$ and $B^{\pm}_{(1,3)}$ indicate those for the nucleon and baryon with its spin-parity $(1,3)^{\pm}/2$, respectively. $F^{\mu\nu}_V$ stands for the anti-symmetric field strength tensor for the massive vector meson. Here, we define the notation $\Gamma^{+,-}=(\gamma_5,\bm{1}_{4\times4})$ corresponding to the parities. The invariant amplitudes can be evaluated straightforwardly, using the effective Lagrangians given above. 

Here, we have an issue that the phase factors between the scattering amplitudes are not uniquely determined. The total scattering amplitude can be written with the phase factors in general as follows:
%EQUATION>>>
\begin{equation}
\label{eq:PHASE}
\mathcal{M}_\mathrm{total}
=e^{i\psi_{t_1,K^*(1680)}}\mathcal{M}_{t_1,K^*(1680)}
+e^{i\psi_{t_4,P^+_s}}\mathcal{M}_{t_4,P^+_s}+e^{i\psi_{t,\mathrm{BKG}}}\mathcal{M}_{t,\mathrm{BKG}}
+e^{i\psi_{u,\mathrm{BKG}}}\mathcal{M}_{u,\mathrm{BKG}},
\end{equation}
%EQUAITON>>>
where the first and second terms in the r.h.s. of Eq.~(\ref{eq:PHASE}) provide peak structures from $K^*(1680)$ and $P^+_s$, whereas the third and fourth ones are for the $t$- and $u$-channel backgrounds (BKG), respectively. The corresponding phase angles are given by $\psi$ here. The phase angles will be determined to fit available experimental data in the next Section.

In order to take the spatial extension of the hadrons into account, the following phenomenological form factors are employed as follows:
%EQUATION>>>
\begin{equation}
\label{eq:FF}
F(q^2;M)=\left[\frac{\Lambda^4}{\Lambda^4+(q^2-M^2)^2}\right]^{1/2}.
\end{equation}
%EQUAITON>>>
As shown in Eq.~(\ref{eq:FF}), because the form factors are functions of the off-shell momenta of the intermediate hadrons, two form factors are multiplied to a bare scattering amplitude in general: $i\mathcal{M}^K_{t_1}\to i\mathcal{M}^K_{t_1}F(q^2_{1-3},M_K)F(q^2_{5-2},M_\phi)$, where $q_{i\pm j}\equiv k_i\pm k_j$, for instance.  The cutoff mass is determined to reproduce relevant experimental data, and will be discussed in the next Section in detail. Note that $g_{KK\phi}$, $g_{\phi NN}$, and $g_{KN\Lambda}$ come from the Nijmegen soft-core potential model~\cite{Rijken:1998yy}, while $g_{\phi P^+_sP^+_s}$ and $g_{KN\Theta^{++}_{27}}$ are estimated theoretically in Refs.~\cite{Khemchandani:2011et} and \cite{Wu:2003xc}, respectively. However, $g_{KN\Theta^{++}_{27}}$ will be treated as a free parameter in the actual calculations, since the maximum value $2.06$ is too large to describe the data. $g_{KN\Lambda(1405)}$ and $g_{KN\Lambda(1520)}$ are determined using the interaction Lagrangians in Eq.~(\ref{eq:LAG}) and experimental data~\cite{Tanabashi:2018oca}. Finally, $g_{KK^*\phi}$ will be determined by fitting with the experimental data, since we do not know its branching ratio to $K^+\phi$ exactly, although its full decay width reads $322$ MeV from the Review of Particle Physics~\cite{Tanabashi:2018oca}. Relevant coupling constants and full decay widths for the present numerical calculations are listed in Table~\ref{TAB0}.
%TABLE>>>
\begin{table}[b]
\begin{tabular}{c|c|c|c|c|c|c|c}
$g_{KK\phi}$~\cite{Tanabashi:2018oca}
&$g_{\phi NN}$~\cite{Rijken:1998yy}
&$g_{KK^*(1680)\phi}$
&$g_{\phi P^+_sP^+_s}$~\cite{Khemchandani:2011et}
&$g_{KN\Theta^{++}_{27}}$~\cite{Wu:2003xc}
&$g_{KN\Lambda}$~\cite{Rijken:1998yy}
&$g_{KN\Lambda(1405)}$~\cite{Tanabashi:2018oca}
&$g_{KN\Lambda(1520)}$~\cite{Tanabashi:2018oca}\\
\hline
$4.75$
&$-1.47$
&$1.6$\,(fit)
&$0.14+0.2i$
&$\lesssim2.06$
&$-13.4$
&$1.51$
&$10.5$\\
\hline
\hline
\multicolumn{2}{c|}{$\Gamma_\phi$~\cite{Tanabashi:2018oca}}
&$\Gamma_{K^*(1680)}$~\cite{Tanabashi:2018oca}
&$\Gamma_{P^+_s}$~\cite{Khemchandani:2011et}
&$\Gamma_{\Theta^{++}_{27}}$~\cite{Wu:2003xc}
&$-$
&$\Gamma_{\Lambda(1405)}$~\cite{Tanabashi:2018oca}
&$\Gamma_{\Lambda(1520)}$~\cite{Tanabashi:2018oca}\\
\hline
\multicolumn{2}{c|}{$4.249$ MeV}
&$322$ MeV
&$14$ MeV
&$\lesssim43$ MeV
&$-$
&$50.5$ MeV
&$15$ MeV\\
\end{tabular}
\caption{Relevant coupling constants and full decay widths.}
\label{TAB0}
\end{table}
%TABLE<<<

%-------------------------------------------------
\section{Numerical results and Discussions}
\begin{figure}[t]
\includegraphics[width=8.5cm]{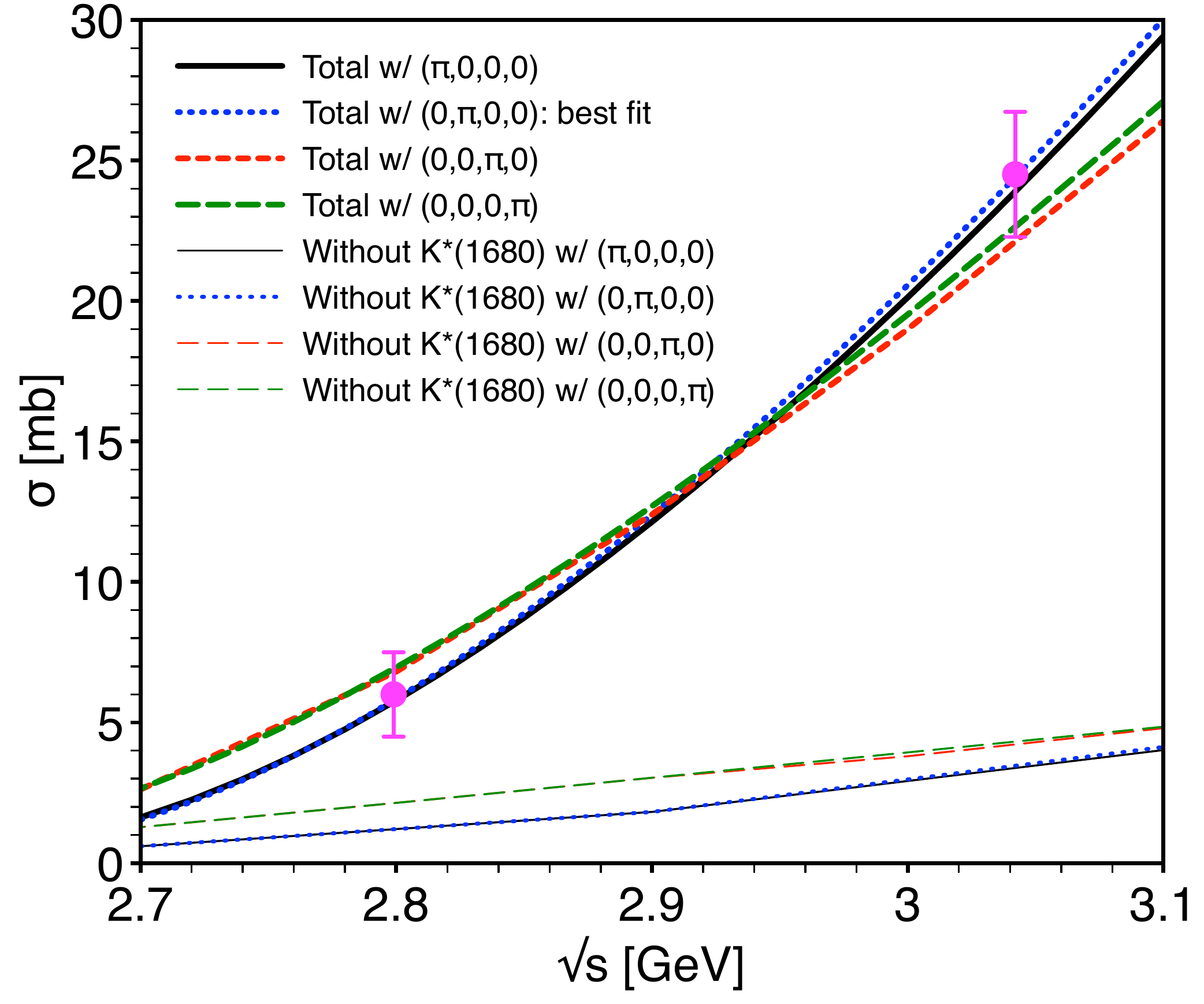}
\caption{(Color online) Total cross sections for $K^+p\to K^+\phi\,p$. The experimental data are taken from Refs~\cite{Bishop:1970at,Moser:1977mp,Flaminio:1983fw}. The total results are given by the thick lines with $(\psi_{t_1,K^*(1680)},\psi_{t_4,P^+_s}\psi_{t,\mathrm{BKG}},\psi_{u,\mathrm{BKG}})=(\pi,0,0,0)$ (solid), $(0,\pi,0,0)$ (dotted), $(0,0,\pi,0)$ (dashed), and $(0,0,0,\pi)$ (long-dashed). The thin lines are for the cases without $K^*(1680)$.} 
\label{TCS}
\end{figure}

%-------------------------------------------------
In this Section, we will provide numerical results and corresponding discussions in detail for the reaction process of $K^+p\to K^+\phi\,p$, focusing on the exotic pentaquark bound-state $P^+_s(2071,3/2^-)$. First, it is necessary to decide the cutoff masses for the form factors in Eq.~(\ref{eq:FF}). For example, the same form factor scheme needs a cutoff mass $\Lambda=(400\sim500)$ MeV to reproduce the total and differential cross sections of the $K^-p$ elastic-scattering process, using the conventional tree-level Feynman diagrams without resonances. For brevity, we assume that the cutoff masses are the same for all the scattering amplitudes and ignore the $\Theta^{++}_{27}$ contribution for a while. {We perform a fitting process by tuning the cutoff mass,  $g_{\phi KK^*(1680)}$, and phase angles in Eq.~(\ref{eq:PHASE}) simultaneously, and the numerical results for the total cross section of $K^+p\to K^+\phi\,p$ are given in Fig.~\ref{TCS}. The experimental data are taken from Refs.~\cite{Flaminio:1983fw,Bishop:1970at,Moser:1977mp}. By fitting with the data, we first determine the cutoff mass and $g_{\phi KK^*(1680)}$, resulting in $480$ MeV and about $1.6$, respectively. Note that the obtained cutoff mass is not much different from that for the $K^-p$ elastic scattering within the same theoretical framework.} 

{As for the phase angles in Eq.~(\ref{eq:PHASE}), we explored various combinations of the angles. Among them, we show the most reliable fit results with $(\psi_{t_1,K^*(1680)},\psi_{t_4,P^+_s}\psi_{t,\mathrm{BKG}},\psi_{u,\mathrm{BKG}})=(\pi,0,0,0)$ (solid), $(0,\pi,0,0)$ (dotted), $(0,0,\pi,0)$ (dashed), and $(0,0,0,\pi)$ (long-dashed). We also verified that other possible combinations, such as  $(\pi,\pi,0,0)$ for instance, are hard to reproduce the cross sections. Although the solid and dotted curves describe the data qualitatively very well, it turns out that the parameter set $(\pi,0,0,0)$ destroys the $P^+_s$ peak structure in the Dalitz plot, due to strong destructive interference between $P^+_s$ and other contributions. Hence, we choose the set $(0,\pi,0,0)$ as our best fit result. Moreover, by seeing the curves without the $K^*(1680)$ contribution (thin lines), we can conclude that the monotonic increasing behavior of the total cross section with respect to $\sqrt{s}$ is reproduced by the $K^*(1680)$ contribution. Hereafter, we will show the numerical results only for the $(0,\pi,0,0)$ case as our final results.}
%FIGURE
%FIGURE

\begin{figure}[t]
%\end{tabular}
\begin{tabular}{cc}
\includegraphics[width=8.5cm]{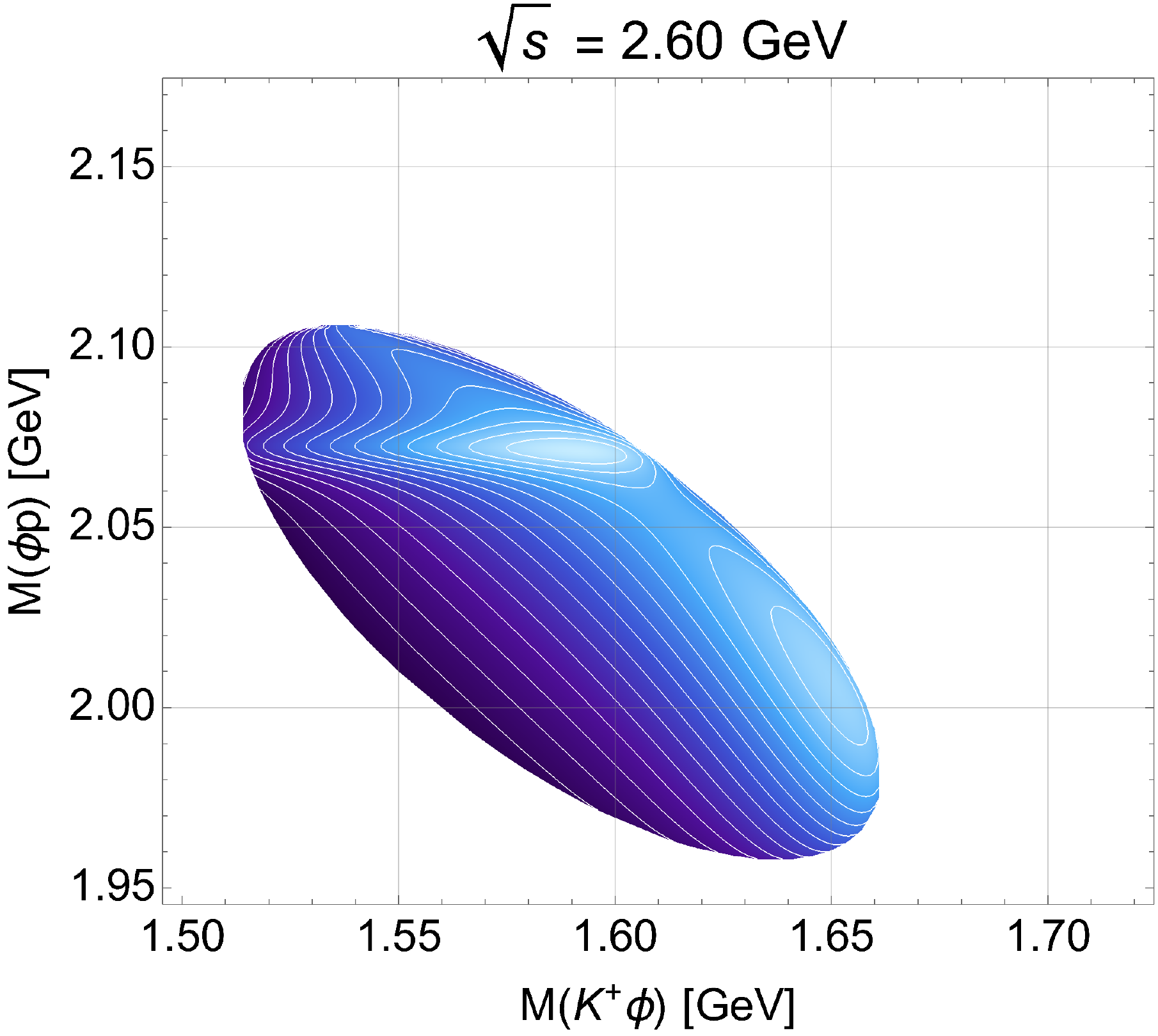}
\includegraphics[width=8.5cm]{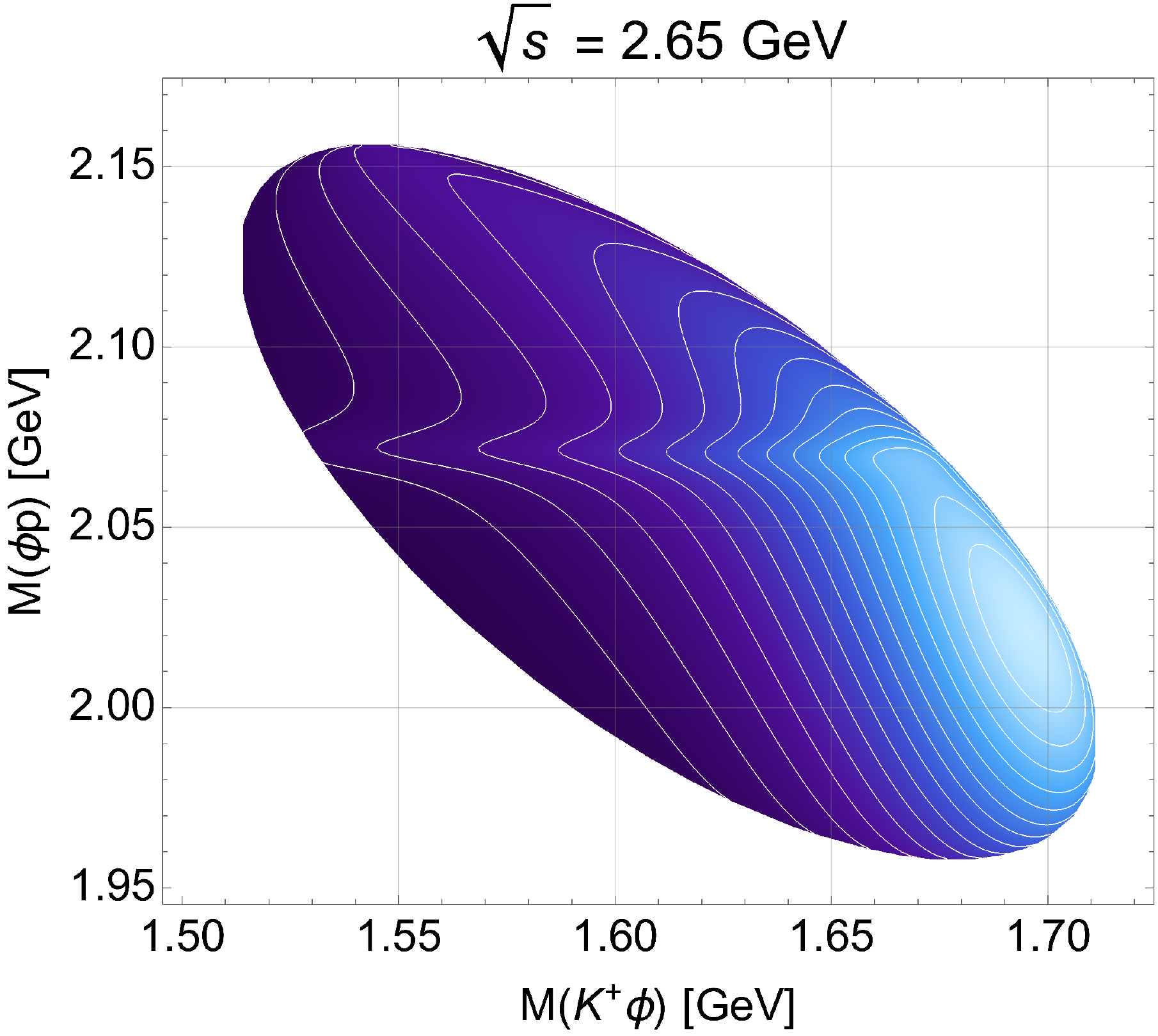}
\end{tabular}
\caption{(Color online) Dalitz plots for $K^+p\to K^+\phi\,p$ for $\sqrt{s}=2.60$ GeV (left) and $2.65$ GeV (right).}
\label{DAL}
\end{figure}

In Fig.~\ref{DAL}, we depict the Dalitz plots for $K^+p\to K^+\phi\,p$ for the various cm energies $\sqrt{s}=2.60$ GeV (left) and $2.65$ GeV (right). Note that the band structures for $P^+_s$ are clearly shown. At $\sqrt{s}=2.60$ GeV, which is about $150$ MeV above the threshold, the structure of the Dalitz plot is simple, i.e., the sum of the background and $P^+_s$ contributions. As the energy increases, the $K^*(1680)$ contribution becomes stronger as shown in the right corner of the Dalitz plot for $\sqrt{s}=2.65$ GeV and starts to interference between $P^+_s$. As expected from the total cross section in Fig.~\ref{TCS}, as the energy gets higher, the $K^*(1680)$ contribution plays a dominant role to reproduce the strength of the cross sections.
%FIGURE
%FIGURE

\begin{figure}[t]
\begin{tabular}{cc}
\includegraphics[width=8.5cm]{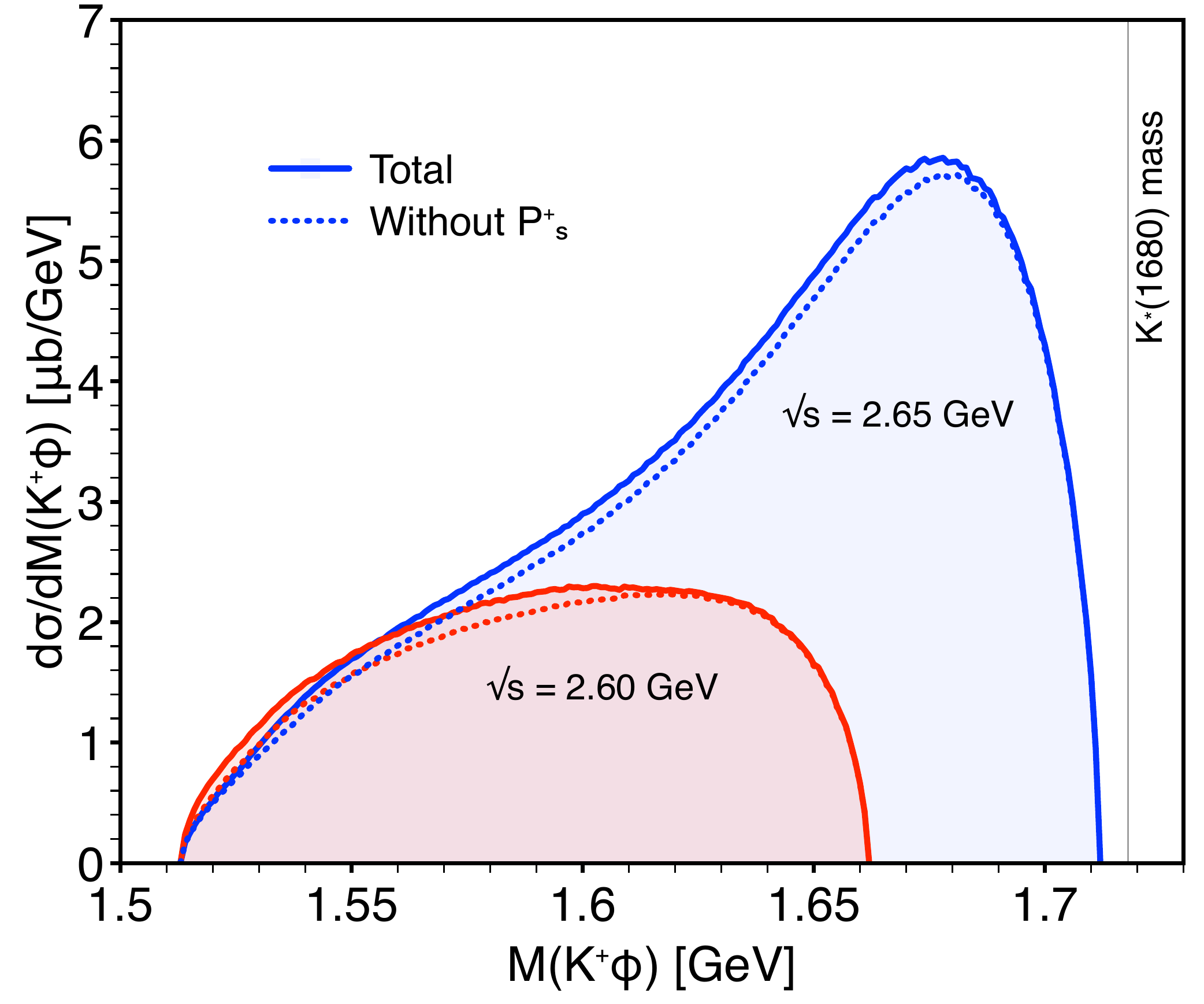}
\includegraphics[width=8.5cm]{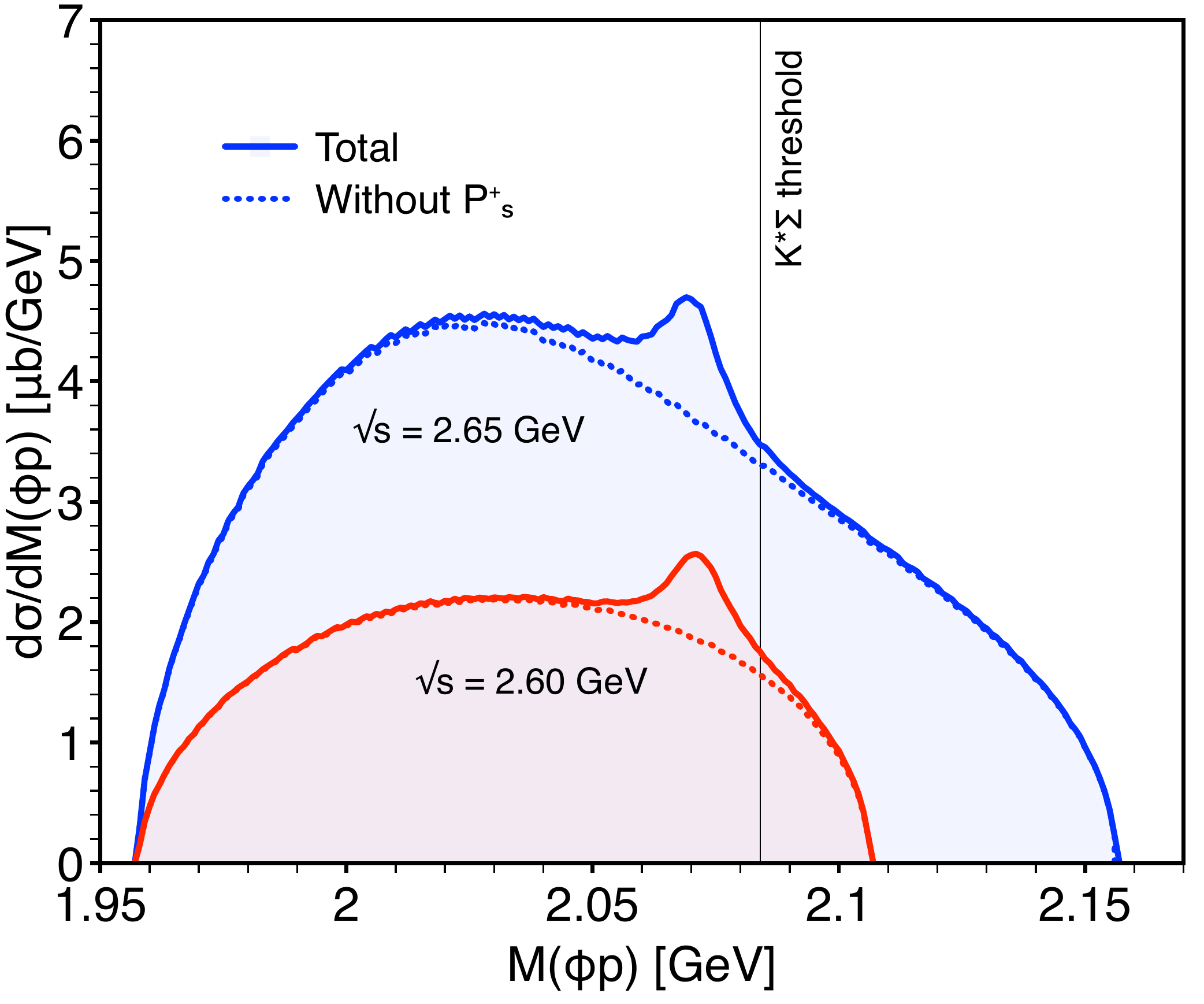}
\end{tabular}
\caption{(Color online) Invariant-mass plots $d\sigma/dM(K^+\phi)$ (left) and $d\sigma/dM(\phi p)$ (right) for $\sqrt{s}=(2.60,2.65)$ GeV. The solid and dotted lines indicate the results with and without $P^+_s$, respectively. }
\label{INV}
\end{figure}

In order to understand the details of the $P^+_s$ band structure and the effects of $K^*(1680)$, we show the $K^+\phi$ (left) and $\phi\,p$ (right) invariant-mass plots for $K^+p\to K^+\phi\,p$ for $\sqrt{s}=(2.60,2.65)$ GeV in Fig.~\ref{INV}. The solid and dotted lines indicate the results with and without $P^+_s$, respectively. {In the left panel for $d\sigma/dM(K^+\phi)$, we observe that the $P^+_s$ contribution does not make significant contribution here.} In the right panel of Fig.~\ref{INV}, we show the numerical results for $d\sigma/dM(\phi p)$ in the same manner with that of the left one. The vertical solid line denotes the $K^*\Sigma$ threshold. The peak structure of $P^+_s$ is clearly observed below the $K^*(892)$-$\Sigma$ threshold, and it gives a signal-to-background ratio $\sim1.7\,\%$ for $\sqrt{s}=2.65$ GeV for instance. {We also verified that the $K^*(1680)$ contribution is responsible for the stiff increase of the cross sections with respect to $\sqrt{s}$ for the both invariant-mass plots.}
%FIGURE
%FIGURE

\begin{figure}[t]
\begin{tabular}{cc}
\includegraphics[width=8.5cm]{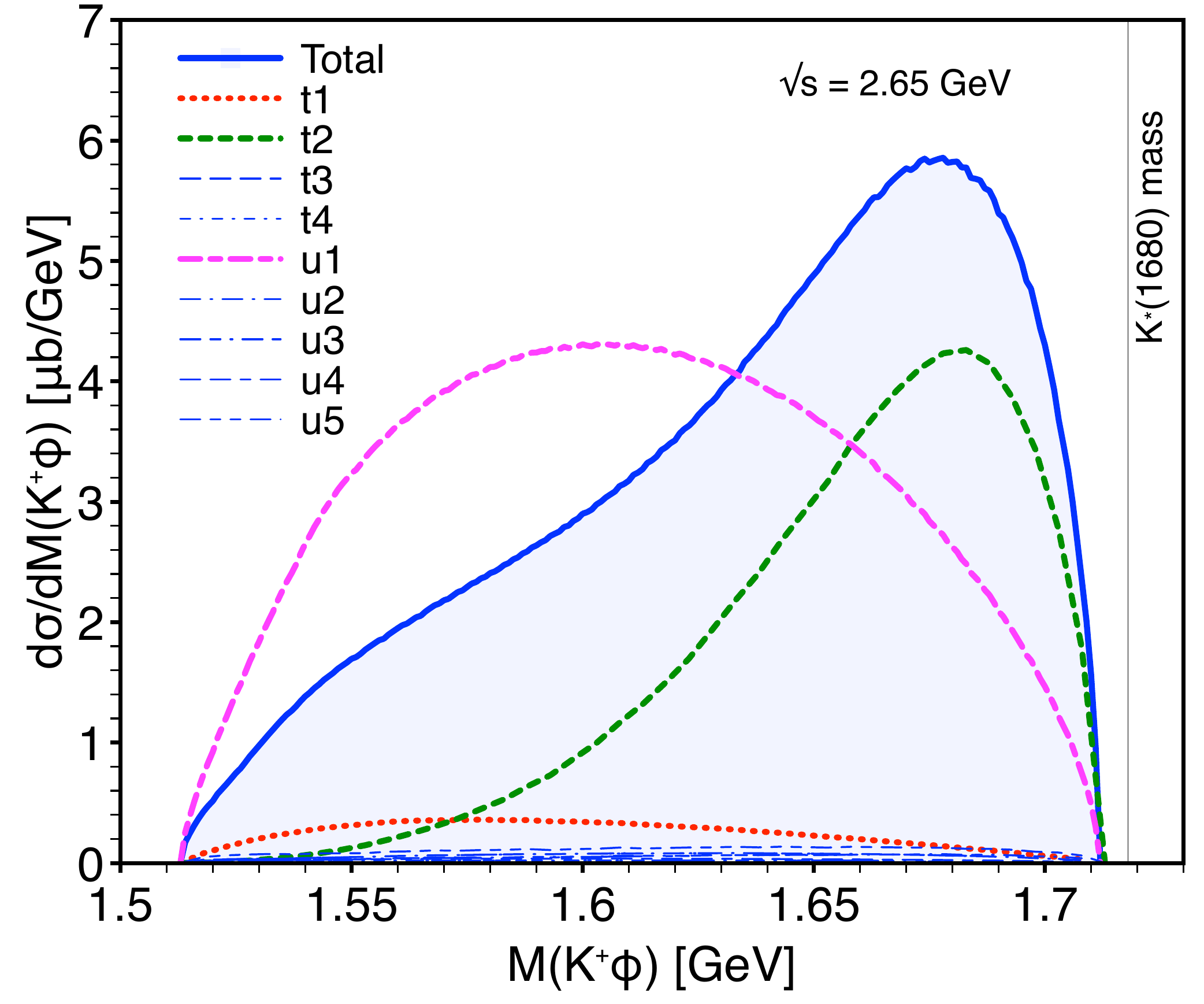}
\includegraphics[width=8.5cm]{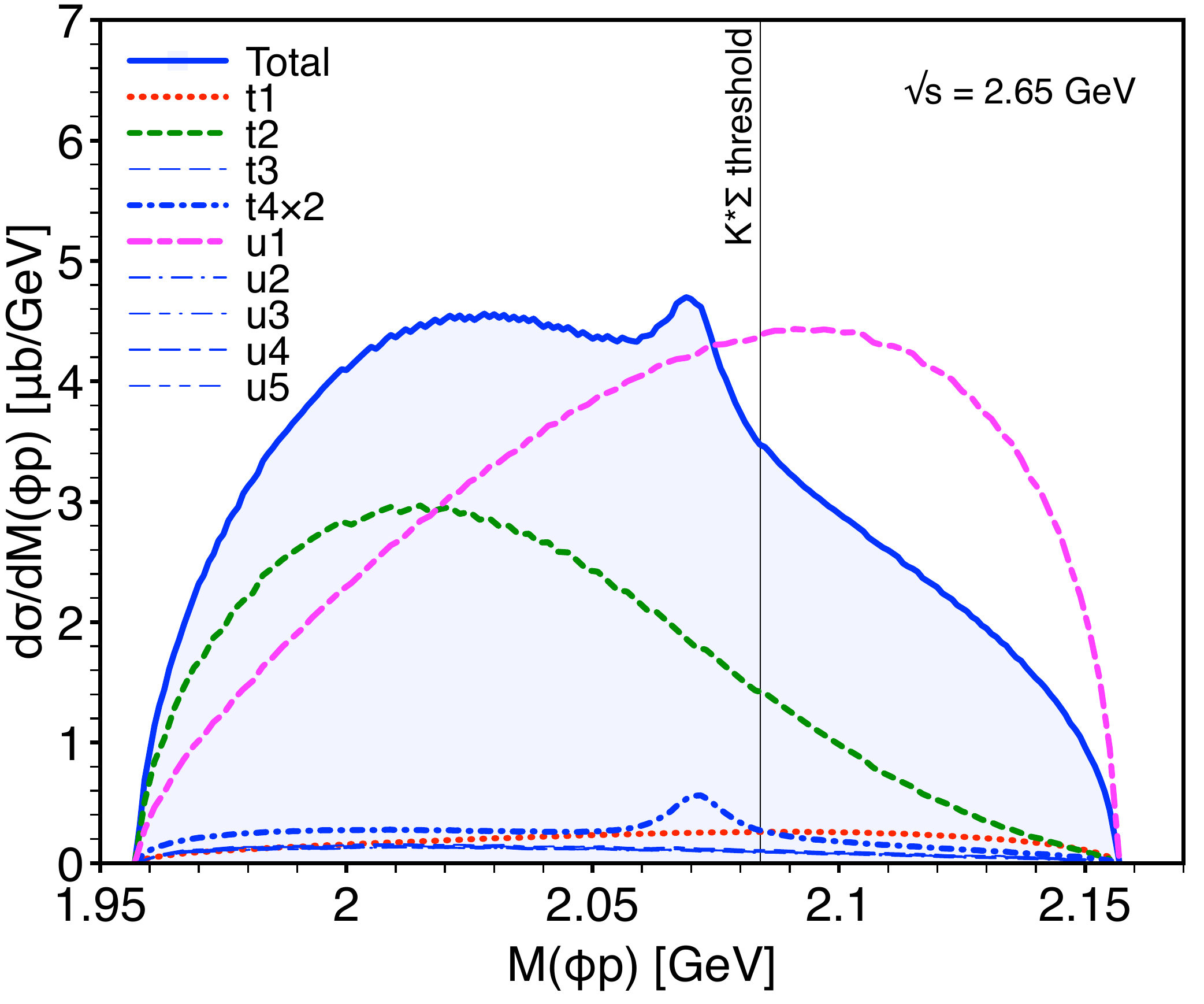}
\end{tabular}
\caption{(Color online) Each contribution for the invariant-mass plots $d\sigma/dM(K^+\phi)$ (left) and $d\sigma/dM(\phi p)$ for $\sqrt{s}=2.65$ GeV (right). Each contribution is given in different line styles as indicated.}
\label{INVEACH}
\end{figure}

In order to examine the interference patterns between the relevant contributions, in Fig.~\ref{INVEACH}, we draw each contribution separately for $d\sigma/dM(K^+\phi)$ (left) and $d\sigma/dM(\phi p)$ (right) at $\sqrt{s}=2.65$ GeV. From the left panel of the figure, we note that the diagram $(t_2)$ with $K^*(1690)$ (dashed) and diagram $(u_1)$ with $\Lambda(1115)$ (dash-long dashed) of Figure~\ref{FIG0} provide the dominant contributions to the differential cross sections. {We note that there appears an abvious destructive interference between those two contributions.} Other diagrams are contributing only small or negligible portion to the total differential cross section. Note that the $P^+_s$ contribution is almost unseen here. In the right panel of Fig.~\ref{INVEACH}, we plot $d\sigma/dM(\phi p)$ in the same manner. Again, the shape of the total curve is almost made from the diagrams $(t_2)$ and $(u_1)$ being similar to $d\sigma/dM(K^+\phi)$, while the diagram $(t_4)$ gives the $P^+_s$ peak, whose strength is enhanced by the constructive interference with other contributions.  
%FIGURE
%FIGURE

\begin{figure}[t]
\begin{tabular}{cc}
\includegraphics[width=8cm]{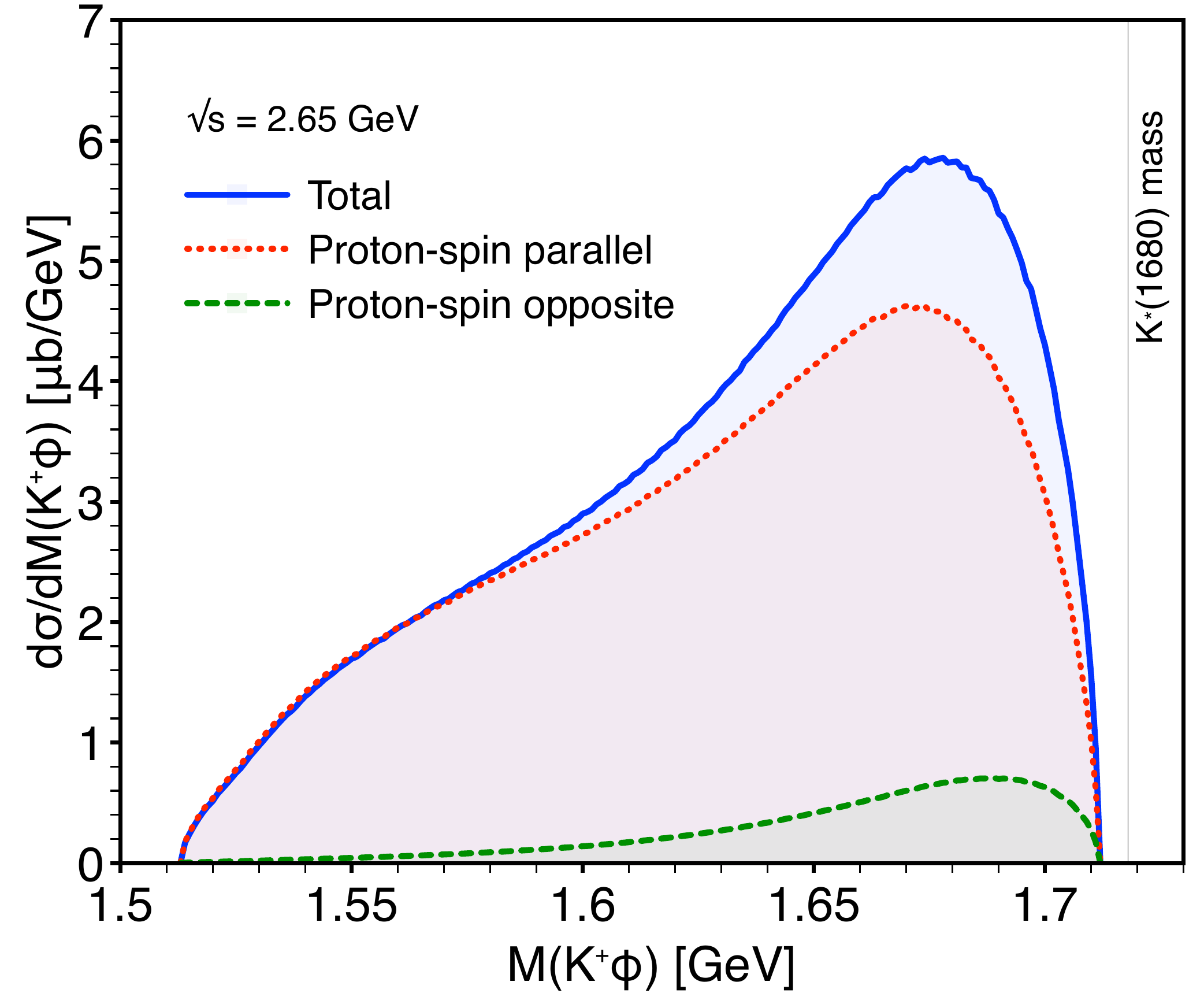}
\includegraphics[width=8cm]{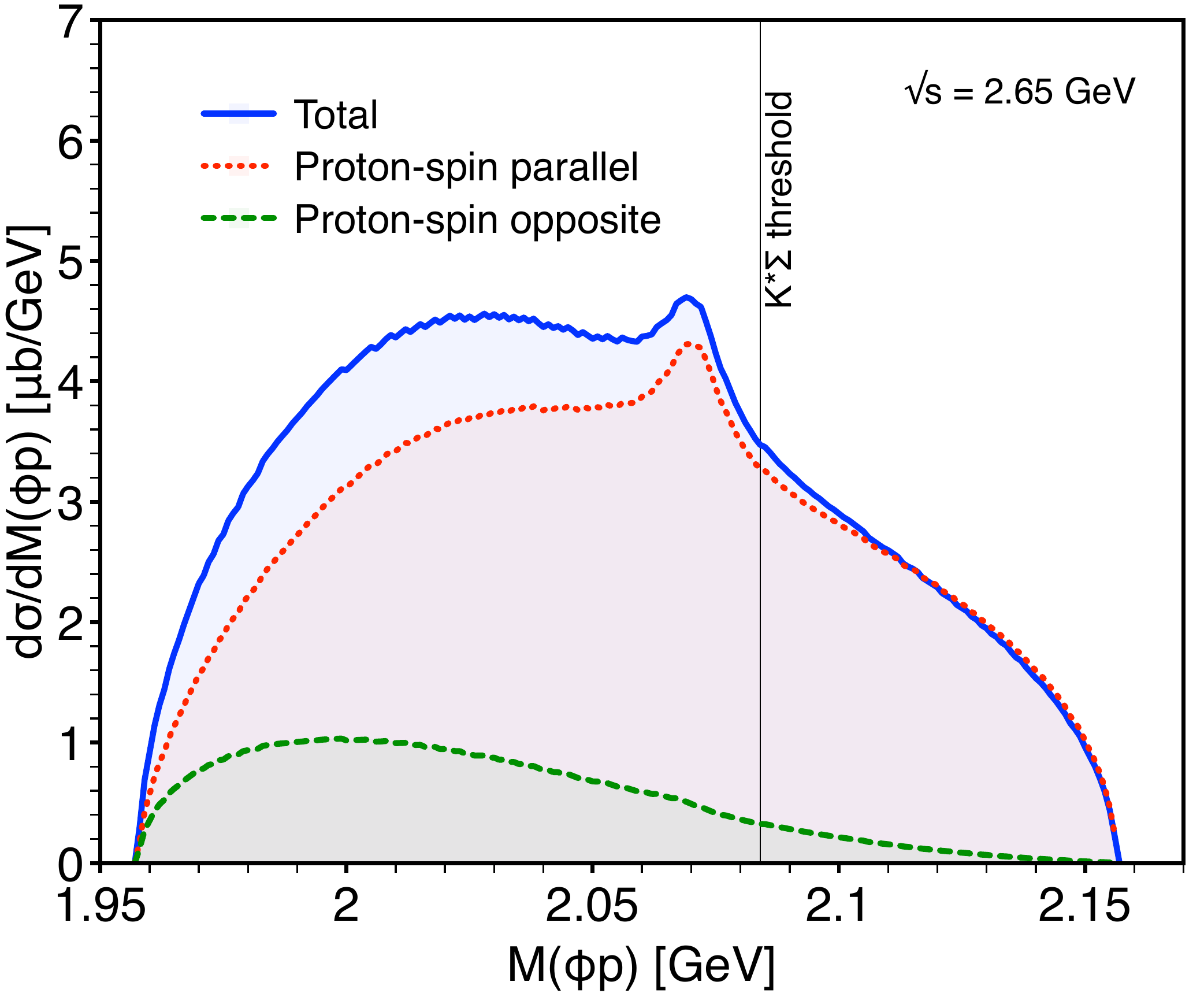}
\end{tabular}
\caption{(Color online) Polarized invariant-mass plots $d\sigma/dM(K^+\phi)$ (left) and $d\sigma/dM(\phi p)$ (right) for $\sqrt{s}=2.65$ GeV. The dotted and dashed lines denote the proton-spin parallel and opposite combinations, respectively, as defined in Eq.~(\ref{eq:SC}). }
\label{INVPOL}
\end{figure}

Now we are in a position to discuss a method to reduce the background contributions to enhance those particle signals that we are interested in. Here, to make the discussion simple, we consider the most important contributions, i.e., diagrams $(t_2)$ for $K^*(1680)$ , $(t_4)$ for $P^+_s$, and $(u_1)$ for a pure background as understood in Fig.~\ref{INVEACH}. For this purpose, we introduce different target initial-state ($i$) and final-state ($f$) proton-spin combinations, i.e., \textit{parallel} or \textit{opposite} to each other:
%EQUATION>>>
\begin{equation}
\label{eq:SC}
\sigma_\mathrm{parallel}\equiv\sigma(\uparrow\uparrow)+\sigma(\downarrow\downarrow),\,\,\,\
\sigma_\mathrm{opposite}\equiv\sigma(\uparrow\downarrow)+\sigma(\downarrow\uparrow),
\end{equation}
%EQUAITON>>>
where we define these cross sections by $\sigma(S^i_zS^f_z)$. By a simple Clebsch-Gordan coefficient analyses, as for the the diagrams $(t_{2,4})$, $\sigma^{t_{2,4}}_\mathrm{parallel}$ and $\sigma^{t_{2,4}}_\mathrm{opposite}$ are both finite, whereas $\sigma^{u_1}_\mathrm{parallel}$ is finite but $\sigma^{u_1}_\mathrm{opposite}\sim0$, since the spin-$0$ PS mesons couple to the spin-$1/2$ ground-state baryons in the diagram $(u_1)$. Hence, by making the the proton spins opposite to each other, the largest background contribution from the diagram $(u_1)$ is suppressed considerably. In Fig.~\ref{INVPOL}, we show the numerical results for the differential cross sections of Eq.~(\ref{eq:SC}) in the same manner with that of Fig.~\ref{INVEACH}. As shown in the left panel of the figure for $d\sigma/dM(K^+\phi)$, the background nearly disappears for the opposite case as expected, and the $K^*(1680)$ contribution becomes obvious. In the right panel for $d\sigma/dM(\phi p)$, the situation is similar but we observe the dubious $P^+_s$ peak as well for the opposite case. Thus, we can conclude that the $K^*(1680)$ contribution dominates the cross section for the opposite case. {Hence, it is rather useful to investigate the vector kaon properties via experiments by polarizing the proton spins.} 
%FIGURE
%FIGURE

\begin{figure}[t]
\begin{tabular}{cc}
\includegraphics[width=8cm]{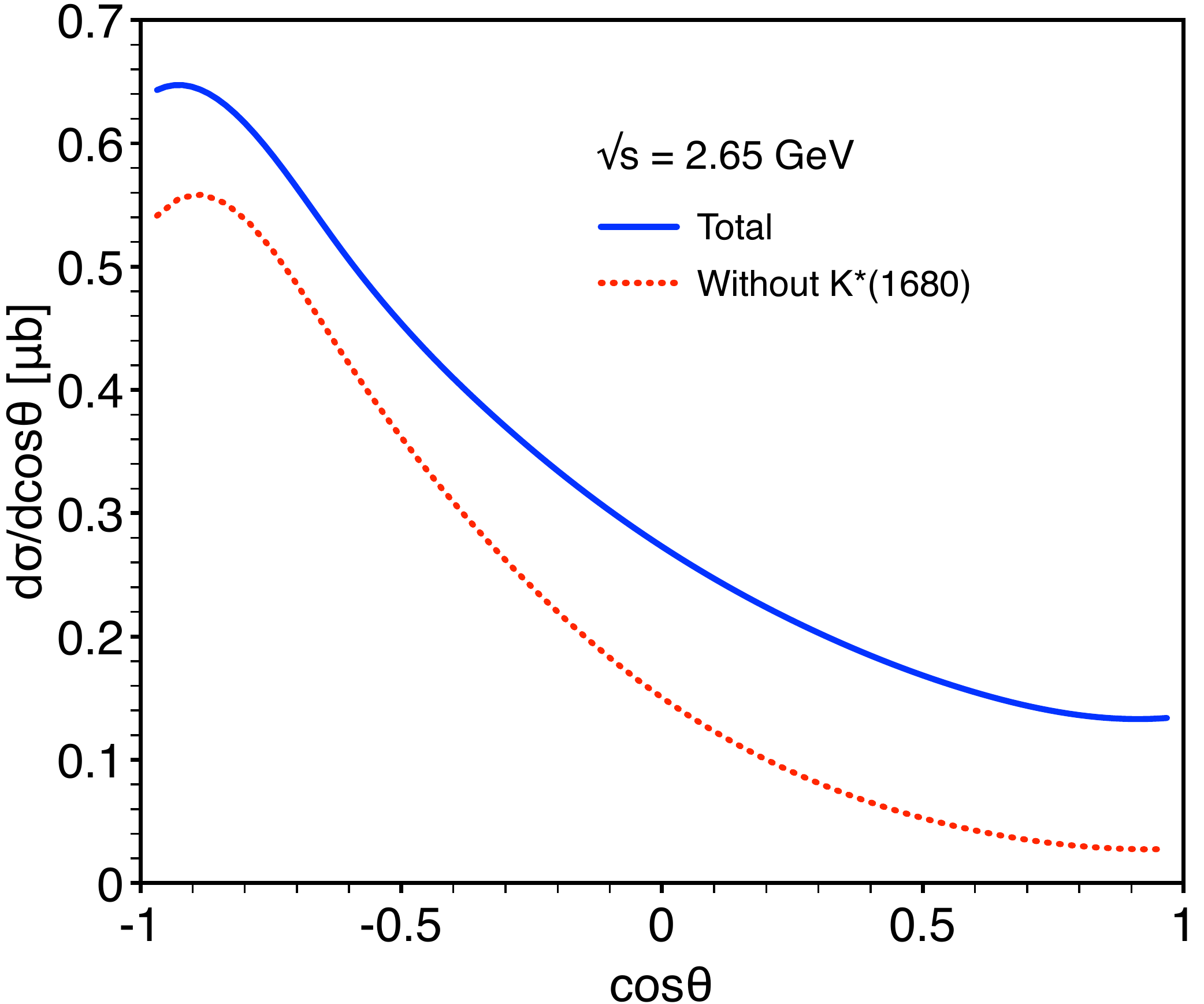}
\includegraphics[width=8.5cm]{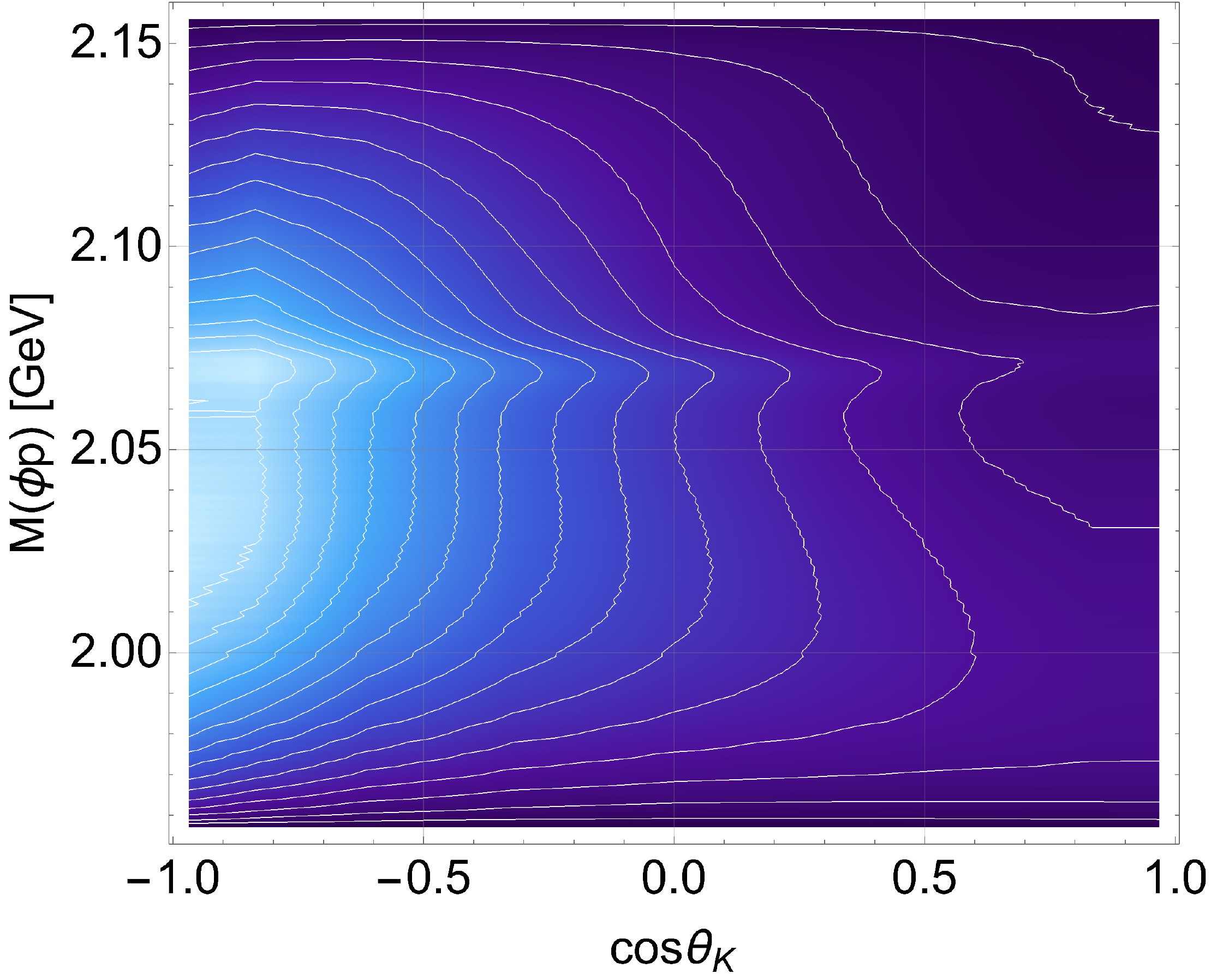}
\end{tabular}
\caption{(Color online) Left: Differential cross section for the angular dependence $d\sigma_{K^+p\to K^+\phi\,p}/d\cos\theta$ for the outgoing-$K^+$ angle $\theta$ in the cm frame at $\sqrt{s}=2.65$ GeV. The solid and dotted lines indicate the cases with and without $K^*(1680)$, respectively. Right: Double-differential cross section $d^2\sigma_{K^+p\to K^+\phi\,p}/dM(\phi p)\,d\cos\theta$ at $\sqrt{s}=2.65$ GeV.}
\label{DCS}
\end{figure}

The numerical results for the angular distribution of the cross section $d\sigma_{K^+p\to K^+\phi\,p}/d\cos\theta$ for $\sqrt{s}=2.65$ GeV is given in the left panel fo Fig.~\ref{DCS}. The angle $\theta$ is defined by that of the outgoing kaon in the cm frame with respect to the $+z$ beam direction. We observe the considerable enhancement in the backward scattering region for the both cases with (solid) and without (dotted) $K^*(1680)$. This behavior can be easily understood from the amplitude for the diagram $(u_1)$, which is the largest contribution of the present reaction process. From the $u$-channel propagator in the amplitude in the cm frame
%EQUATION>>>
\begin{equation}
\label{eq:U1}
i\mathcal{M}_{u_1}\propto
\left[\left(M^2_{\Lambda^{(*)}}-M^2_N-M^2_K\right)+2E_{p_i}E_{K^+_f}-2\left(\vec{k}_{p_i}\cdot \vec{k}_{K^+_f}\right)\right]^{-1},
\end{equation}
%EQUAITON>>>
we find that, when the three momenta of the target proton $\vec{k}_{p_i}$, which is in the $-z$ direction, and the outgoing kaon $ \vec{k}_{K^+_f}$ are parallel in the cm frame, the amplitude becomes maximized. {By comparing the two curves, we can deduce that the $K^*(1680)$ contribution gives a slightly forward-enhancing but rather flat angular dependence, as understood by the diagrams ($t_{1,2}$).} In the right panel of Fig.~\ref{DCS}, we show the double differential cross section $d^2\sigma_{K^+p\to K^+\phi\,p}/dM(\phi\,p)d\cos\theta$ at $\sqrt{s}=2.7$ GeV. There, we clearly see that the $P^+_s$ peak manifests itself in the backward-scattering region $\cos\theta\lesssim-0.5$. 
%FIGURE
%FIGURE

\begin{figure}[t]
\begin{tabular}{cc}
\includegraphics[width=8.5cm]{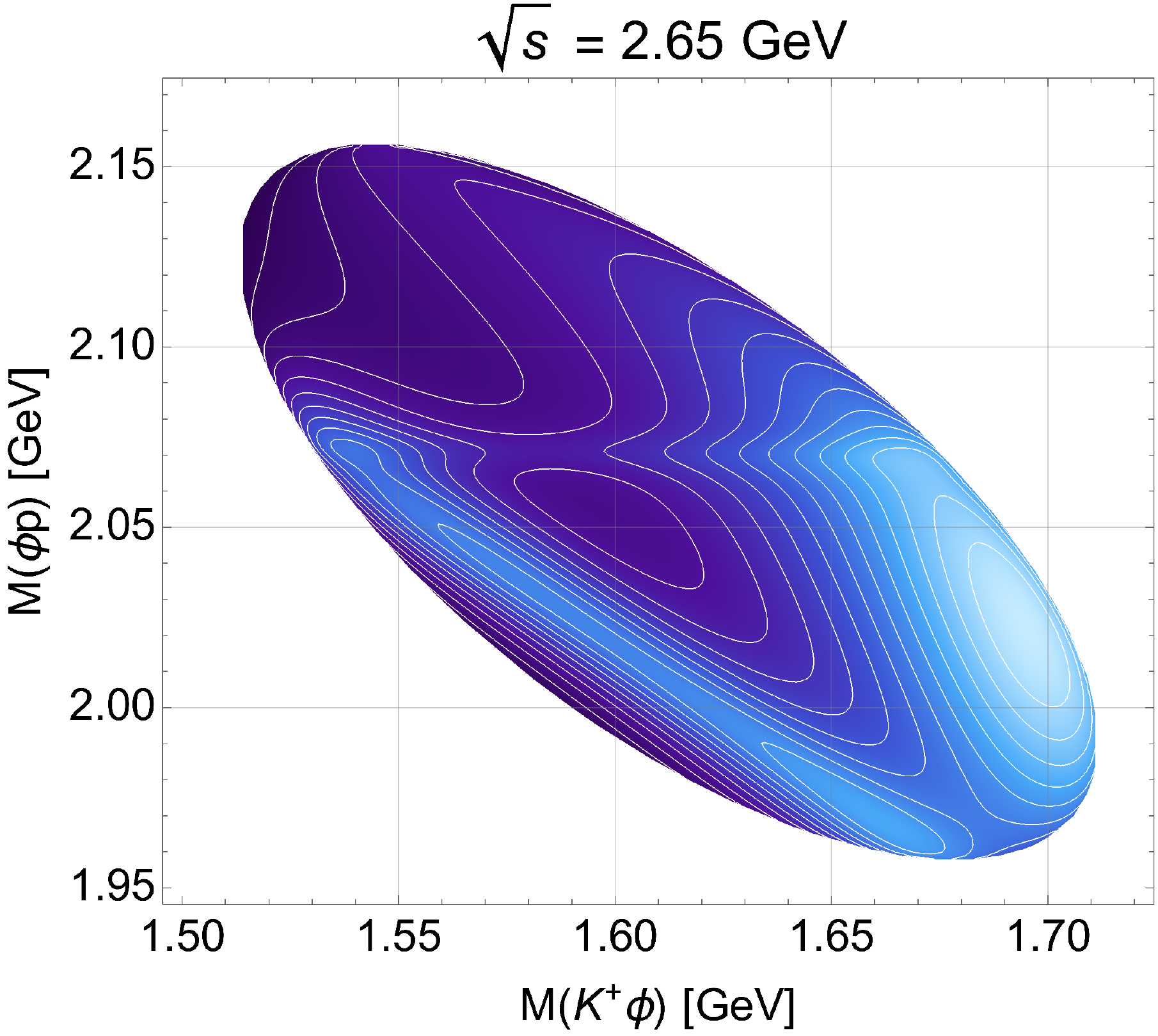}
\includegraphics[width=8.5cm]{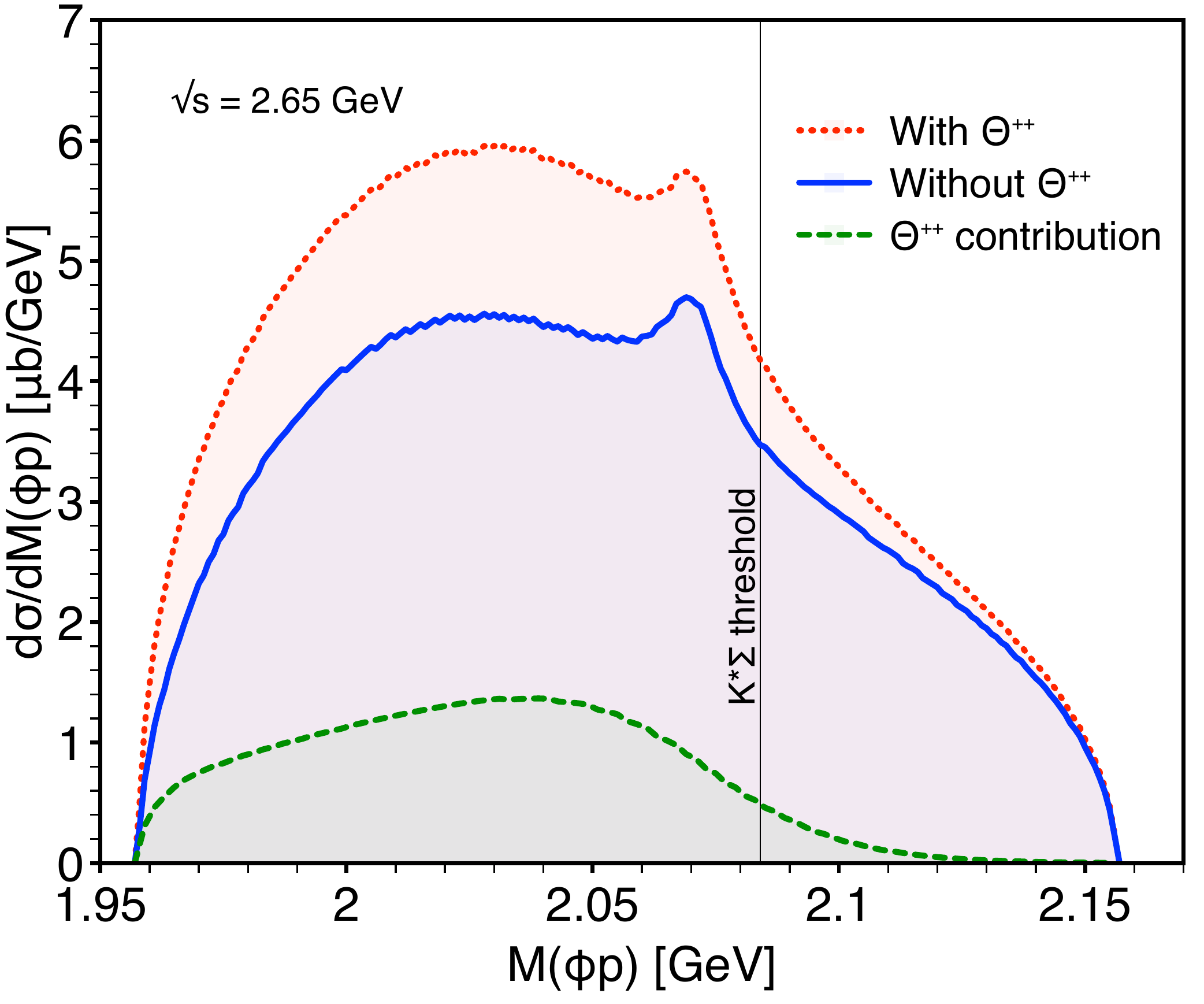}
\end{tabular}
\caption{(Color online) Left: Dalitz plots for $K^+p\to K^+\phi\,p$ for $\sqrt{s}=2.65$ GeV with the $\Theta^{++}_{27}$ contribution. Right: Differential cross section $d\sigma/dM(\phi p)$ for $\sqrt{s}=2.65$ GeV with (dotted) and without (sloid) the pentaquark contribution. The $\Theta^{++}_{27}$ contribution is given by the dashed line. }
\label{PENTA}
\end{figure}

Finally, we explore the effect from the possible $27$-plet pentaquark $\Theta^{++}_{27}$, appearing in the diagram $(s_1)$ of Fig.~\ref{FIG0}. {We employ the theoretical information for the exotic baryon from the chiral-soliton model~\cite{Wu:2003xc}: $M_{\Theta^{++}_{27}}=1.6$ GeV and $\Gamma_{\Theta^{++}_{27}\to KN}\lesssim43$ MeV. Considering that the pentaquark can be spatially larger than usual ones, one can modify the phenomenological form factor in Eq.~(\ref{eq:FF}), and we introduce more strong form factor as $F(q^2;M) \to F^r(q^2;M)$, where the free parameter $r$ indicates a positive-real constant. Since there have been no experimental data for the pentaquark, we simply choose the value of $r$ to provide a similar strength of the cross section in comparison to those from other non-exotic contributions. As a trial, the value of $r$ is chosen to be $2.5$ and, in the left panel of Fig.~\ref{PENTA}, we depict the numerical results for the Dalitz plot at $\sqrt{s}=2.65$ GeV. There appears a diagonal band for the pentaquark and it interferes with the $K^(1680)$ contribution. In the right panel of figure, the $\phi\,p$-invariant-mass plots are given with (dotted) and without (solid) the $\Theta^{++}_{27}$ contribution. The sole $\Theta^{++}_{27}$ contribution is depicted by the dashed line. By seeing that, we can conclude  that the cross section below $M(\phi p)\lesssim 2.05$ will be enhanced more as the pentaquark contribution gets stronger, remaining the $P^+_s$ peak intact.} 
%FIGURE
%FIGURE

%-------------------------------------------------
\section{Summary}
%-------------------------------------------------
In the present work, we investigated the hidden-strangeness production process via $K^+p\to K^+\phi\,p$. Here, assuming that the hidden-flavor pentaquark molecular bound state is easier to be formed than open-flavor ones, since $P^+_c(4457)$ is the only exotic pentaquark baryon ever observed firmly in $P^+_c[uudc\bar{c}]\to J/\psi[c\bar{c}]\, p$, we considered that $P^+_s[uuds\bar{s}]\to\phi[s\bar{s}]\,p$, in which $P^+_s$ is an exotic molecular bound state of $K^*$ and $\Sigma$, can be observed in this specific production process with higher possibility than other light-flavor baryonic exotics. We employed a simple phenomenological model based on the effective Lagrangian approach and provided the relevant numerical results by making use of presently available experimental and theoretical information for the purpose. We used the theory estimations for the $P^+_s(2071,3/2^-)$ from the hidden-local symmetry arguments. In addition to the background and $P^+_s$ contributions, we also took into account $K^*(1680)$, which decays into $K\phi$, as indicated in the LHC$_b$ experiments. Below, we list up the relevant observations in the present work:
%ITEMIZE
\begin{itemize}
%-------------------------------------------------
\item First we determined the cutoff mass, which is one of the most important model parameters, by fitting the experimental data for the total cross section of $K^+p\to K^+\phi\,p$ for $\sqrt{s}\gtrsim2.7$ GeV. We observed strong enhancements of the cross section with the $K^*(1680)$ contribution as the cm energy increases, due to the higher momentum-dependent nature of the $KK^*\phi$ interaction vertex. From the Dalitz plot analyses, we observed the narrow ($\Gamma=14$ MeV) band structure for $P^+_s$ with the increasing background contributions as the cm energy does, and the $P^+_s$ contribution is constructively interfered with the background. As the energy becomes larger than $\sqrt{s}\approx2.65$ GeV, the $K^*(1680)$ contribution comes into play and starts to dominate the cross section of the process. 
%-------------------------------------------------
\item From the $\phi\,p$-invariant-mass plots, we found an obvious peak structure from $P^+_s$ with the signal-to-background ratio $\approx1.7\%$ at $\sqrt{s}=2.65$ GeV. There were considerable cross section enhancements from the $K^*(1680)$ contribution for $M(\phi p)\lesssim2.05$  GeV. We could observe a peak-like structure from the $K^*(1680)$ contribution as the cm energy increases in the $K^+\phi$-invariant-mass plots. Beside the $P^+_s$ and $K^*(1680)$ contributions, the scattering amplitude ($u_1$) with the $\Lambda(1115)$-hyperon intermediate states with the $\phi$ meson emitted from the kaon beam were the most largest source to produce the cross section as background. 
%-------------------------------------------------
\item We tried to find a way to reduce the background to enhance the signals of the $K^*(1680)$ and $P^+_s$ by different initial- and final-state proton-spin combinations. When the spins are opposite to each other, the backgrounds are greatly suppressed, since the spin non-flip process only survives in the scattering amplitude. As for the proton-spin opposite case, the $K^*(1680)$ contibution was seen with clarity for $\sqrt{s}\gtrsim2.65$ GeV in the $K^+\phi$-invariant-mass plots. On the contrary, the $P^+_s$ peak was dubious for the proton-spin opposite case in the $\phi\,p$-invariant-mass plots, since $P^+_s$ is strongly interfered with the background contributions constructively. 
%-------------------------------------------------
\item The angular dependences of of the cross sections showed mild backward-scattering enhancements with respect to the outgoing $K^+$ angle in the cm frame, and this behavior was originated from the nature of the $u$-channel propagators in the $K^*(1680)$ and background amplitude, mentioned above, i.e., the scattering amplitude is maximized when three momenta of the incident and scattered particles are opposite to each other in the cm frame. Hence, the signal of $P^+_s$ was also amplified in the backward-scattering region as shown in the numerical results of the double differential cross sections $d^2\sigma_{K^+p\to K^+\phi\,p}/dM(\phi p)\,d\cos\theta$.
%-------------------------------------------------
\item Finally, we examined the possible contribution from the $27$-plet pentaquark $\Theta^{++}_{27}$, being based on the theoretical estimations for it from the chiral soliton model. {Modifying the phenomenological form factor for the spatially larger pentaquark, we observed similar cross-section strengths to those of other contributions. The diagonal band structure from $\Theta^{++}_{27}$ interferes softly with $K^*(1680)$ and $P^+_s$ in the Dalitz plot. The cross section was enhanced for $M(\phi p)\lesssim2.05$ GeV as the pentaquark contribution becomes larger.}
%-------------------------------------------------
\end{itemize}
%ITEMIZE

The numerical results of the present work can be a useful guide to measure the exotic pentaquark molecular bound state $P^+_s$ as well as to extract the information of $K^*(1680)$ possibly in the J-PARC experiments with the high-momentum kaon beam in the future. It is interesting to note that Ref.~\cite{Khemchandani:2011et} also suggested another possible $K^*\Sigma$ bound state with its pole mass $(1977-i22)$ MeV and spin-parity $1/2^-$ in addition to $P^+_s(2071,3/2^-)$, although we did not consider this lower-mass state in the present work, since it exists near the Dalitz-plot boundary $M(\phi p)\approx 1960$ MeV and its wider width. The situation is similar to that there are two bound states $P^+_c(4440)$ and $P^+_c(4457)$ below the $\bar{D}^*\Sigma$ threshold. Nonetheless for the mass differences are significantly different $\Delta M_s\approx100$ MeV and $\Delta M_c\approx20$ MeV, the similarities between the light- and heavy-flavor sectors are very peculiar. More detailed works together with more exotics such as $P^+_s(1977)$ in a specific reaction process and theoretical studies for the structures of the exotics are in progress, and appear elsewhere. 

%-------------------------------------------------
\section*{Acknowledgment}
%-------------------------------------------------
The author is grateful to fruitful discussions with K. P. Khemchandani (Sao Paulo Univ.), J.~K.~Ahn (Korea Univ.), and A.~Hosaka (RCNP). This work was supported by a Research Grant of Pukyong National University (2019). All the Feynman diagrams were generated via https://feynman.aivazis.com/.
%-------------------------------------------------

%(^o^)
\end{document}